# Quasi-invariance of scattering properties of multicellular cyanobacterial aggregates

Chunyang Ma [1, 2 *], Qian Lu [3], Yen Wah Tong [2, 4 *]

[1] *School of Advanced Manufacturing, Nanchang University, 999 Xuefu Avenue, Nanchang 330031, People's Republic of China.*

[2] *NUS Environmental Research Institute, National University of Singapore, 1 Create Way, Create Tower #15-02, Singapore 138602, Singapore.*

[3] *School of Grain Science and Technology, Jiangsu University of Science and Technology, Zhenjiang, 212100, China.*

[4] *Department of Chemical and Biomolecular Engineering, National University of Singapore, 4 Engineering Drive 4, Singapore 117585, Singapore.*

**Abstract**

The radiative/scattering properties of cyanobacterial aggregates are crucial for understanding microalgal cultivation. This study analyzed scattering matrix elements and cross-sections of cyanobacterial aggregates using the discrete dipole approximation (DDA) method. The stochastic random walk approach was adopted to generate a force-biased packing model for multicellular filamentous cyanobacterial aggregates. The effects of shape and size of multicellular cyanobacterial aggregates on their scattering properties were investigated by this work. The possibility of invariance in the scattering properties was explored for cyanobacterial aggregates. The invariance interpretation intuitively represented the radiative property characteristics of the aggregates. The presented results show that the ratios of the matrix elements of cyanobacterial aggregates are nearly shape, size, and wavelength invariant. The extinction and absorption cross-sections (EACSs) per unit volume were shape and approximate size invariance of cyanobacterial aggregates, respectively. The absorption cross-section of aggregates is not merely a volumetric phenomenon for aggregates that exceed a certain size. Furthermore, the absorption cross-sections per unit volume are independent of the volumetric distribution of the microalgae cells. The invariance interpretation presents crucial characteristics of the scattering properties of cyanobacterial aggregates. The existence of invariance greatly improves our understanding of the scattering properties of microalgal aggregates. The scattering properties of microalgal aggregates are the most critical aspects of light propagation in the design, optimization, and operation of photobioreactors.

---

*Corresponding author: cyma@ncu.edu.cn (CY Ma), chetyw@nus.edu.sg (YW Tong).

## 1. Introduction

Microalgae-based biomaterials are promising renewable energy sources that have garnered considerable attention in the pharmaceutical industry [1], environmental protection [2], carbon-neutral biofuels [3] and space food supplements [4]. It has been widely recognized that microalgae are renewable green energy sources, meaning that microalgae biomass has no net $CO_2$ emissions in its usage. This balance is achieved by microalgae-based carbon dioxide ($CO_2$) absorption in culture and $CO_2$ releasing in biomass utilization in downstream industry. Therefore, microalgae biomass is conducive to achieving carbon neutrality objectives [5, 6].

Currently, high cultivation costs of microalgae seriously limit their practical application [7]. Although, we have much knowledge of microalgal photosynthesis and production. Microalgae cultivation is affected by light intensity, nutrient concentration, pH, and temperature [8-10]. The light field in a microalgal suspension is one of the most influential factors. Excessive light supply can lead to photoinhibition, whereas insufficient illumination negatively impacts the photosynthesis of algal cells [11]. The interaction between algal cells and illumination can be represented by the scattering properties of microalgae cell, including the absorption, scattering and extinction cross-sections, scattering phase function (SPF), and Mueller matrix elements. Therefore, the scattering properties are regarded as crucial factors which can impact the light transfer within microalgae cultures.

Admittedly, previous studies measured considerable data related to the microalgal radiation characteristics. Berberoglu and Pilon [12] experimentally measured the EACSs of *Anabaenavariabilis* and *Rhodobacter sphaeroides* at 300-1300 nm wavelengths and the SPF of microalgae at a visible wavelength. Berberoglu et al. [13, 14] also reported the measured radiation characteristics of microalgae, including the absorption and scattering cross-sections, SPF, and size distribution of algae cells for more than seven different strains, such as *Botryococcus braunii*, *Chlorococcum* *littorale*, and *Chlamydomonas reinhardtii*. Accordingly, Heng and Pilon [15] experimentally studied the radiation characteristics of *Nannochloropsis oculata* at different growth times under a batch culture condition. Liu's group studied the time-dependent radiation characteristics of typically green and cyanobacterial microalgae, and proposed a temporal scaling law for describing the optical properties that vary with cultivation time [16-18]. Therefore, research on the microalgal radiative properties has experienced a period from stationary to growth, representing an improved understanding of microalgae growth.

In addition to the progress in experimental measurements, previous studies also analyzed scattering properties of microalgae by conducting theoretical study. The Lorentz-Mie theory is typically employed to estimate the scattering properties of spherical unicellular microalgae cells [19, 20]. The effects of internal or external structure of cell on microalgal scattering properties were studied by Dauchet et al. [21] and Dong et al. [22]. Bhowmik and Pilon [23] further studied whether the optical properties of unicellular photoautotrophic spherical-shaped algae cells can be described by single optical homogeneous appropriate spherical cell or two-layer spherical model using the T-matrix calculation approach. They verified that the homogeneous spherical-shaped framework could accurately predict the EACSs of internally non-uniform microalgae. Heng et al. [24] and Kandilian et al. [25] reported the optical properties of regular and irregular colonies of spherical monomers and their effective particle approximation frameworks. Lee and Pilon [26] studied the absorption and scattering of long and randomly oriented sphere chains to approximate some realistic cyanobacteria.

These studies found that the absorption and scattering cross-section per unit length of chain shaped microalgae could be determined by using the infinitely long cylinder solutions. Ma et al. [27] investigated the spectral radiation characteristics of multicellular cyanobacteria using different cell shapes and equivalent optical models. It was discovered that the volume equivalent homogeneous cylinder model provided good predictions on EACSs cross-sections and polarization-dependent Mueller matrix elements of multi-cellular filamentous cyanobacteria. These results show that the microalgal radiation characteristics could be determined by employing optically homogeneous models for both spherical microalgae and multicellular filamentous cyanobacteria. Hoeniges et al. [28] studied the scattering and absorption characteristics of fractal cell colonies of sphere-shaped *Botryococcus braunii* and the effective coated sphere approximation. However, the scattering properties of multicellular filamentous cyanobacterial aggregates remain an important problem, yet to be studied. This study attempted to explore the spectral scattering properties of cyanobacterial aggregates with different shapes, concisely, the effect of the scattering properties of multicellular long chain-like cyanobacteria aggregates in explaining and understanding microalgal growth.

This study investigated the scattering properties and provides theoretical methods for cyanobacteria aggregates important to comprehend and measure how light interacts with aggregation of photosynthetic microalgae. DDA numerical method was adopted to report the spectral scattering properties of multicellular filamentous cyanobacterial aggregates. The effects of the shape and size of the microalgae aggregates on the scattering matrix elements and cross-sections were studied. To interpret the results, the cross-section per unit volume was introduced to represent the characteristics of the microalgal aggregates. In addition, the internal electric fields within the microalgal aggregates were presented to explain the physical mechanism of the variation in the cross-section. This research will improve understanding of the scattering properties of microalgal aggregates and advance the investigation of the radiation characteristics of multicellular filamentous cyanobacterial aggregates.

## 2. Models of photoautotrophic cyanobacteria aggregates

### 2.1. Aggregates of cyanobacteria

*Nostoc* sp. was purchased from the FACHB's freshwater algae species bank, which is located at Wuhan, China. *Nostoc* sp. was cultivated in a bioreactor of which the working volume was 100 mL. The bioreactor was put into the incubator under the cultivation conditions of 12 hours of darkness followed by 12 hours of light, and it was exposed to fluorescent light bulbs with an illumination of 2000–2500 lx. The temperature of the cultivation environment was maintained at 25 ºC. The BG11 medium was used for the cultivation of *Nostoc sp.*.

Microalgae are photosynthetic microorganisms that grow in natural environments or artificial photobioreactors. Microalgae cultures form nonmotile colonies under certain environmental conditions [29]. For example, Ratcliff et al. [30] demonstrated the multicellular complexity of *C. reinhardtii*, which develops from a single cell. Some microalgae strains secrete exopolysaccharides, which are viscous substances coating the surfaces of microalgal cells and causing their aggregation into colonies [31]. Moreover, microalgae aggregation occurs at high cell concentrations owing to the electrostatic force between the cells [32].

Figure 1 shows the micrographs of *Nostoc sp.* at different magnifications. As shown, the *Nostoc sp.* is 2-4 μm in diameter and more than 40 μm in length, and the cells are approximately oval-shaped. Cyanobacteria are usually multicellular with diverse cell shapes, such as spherical, oval, and cylindrical [26].

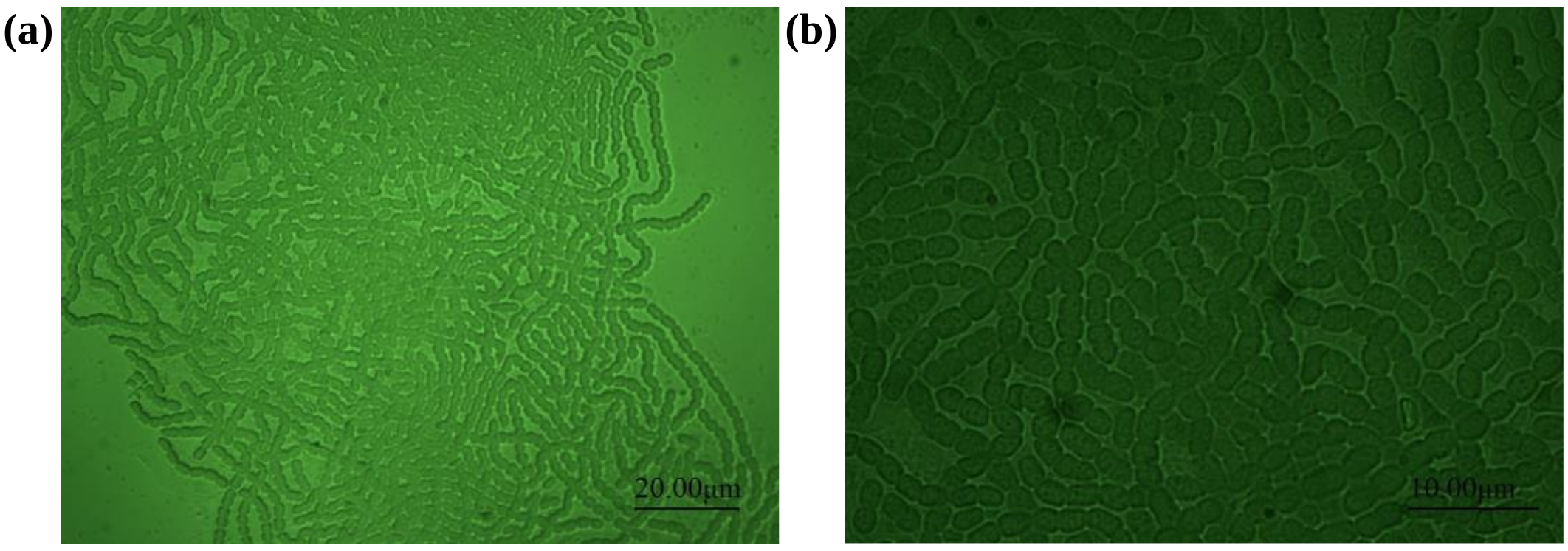

Fig. 1. Micrographs of *Nostoc sp.* (a) 40x, and (b) 100x colonies resembling aggregates.

2.2. Aggregates model establishing

Microalgal cells typically form aggregates during cultivation. The fractal model is typically used to simulate a spherical and unicellular microalgae colony. The cyanobacteria are multicellular species, and their shape is long linear or curved chains. The shape of cyanobacteria is more likely to be curved fibers when details are ignored. We used a stochastic model to generalize the force-biased packing method to multicellular filamentous cyanobacteria in the form of a spherical chain, which can take into account cyanobacteria chain details. This approach is based on using the random walk method, [33]

$$P=\{p_0,\dots,p_n\},\ p_i=(x_i,\mu_i,r_i) \tag{1}$$

where $P$ is the sequence of random walk points. Here it should be understood that the point of random walk is related to its radius and current direction. The initial coordinate $x_0$ is distributed in a cubic uniformly, the radius $r$ and the number of samples $n$ can be fixed for the entire system. The orientation can be generated from the $\beta$ orientation distributions [34]

$$p(\theta,\phi\mid\beta)=\frac{\beta\sin\theta}{4\pi\left[1+\left(\beta^2-1\right)\cos^2\theta\right]^{3/2}} \tag{2}$$

where $(\theta,\phi)$ are the polar coordinates of the orientation $\mu_0$. The global parameter $\beta$ describes the features of orientation distributions, and $\beta=1$ was selected for isotropic orientation distribution here. The bending characteristics of cyanobacterial aggregates are indirectly controlled by two parameters, $\kappa_1$ and $\kappa_2$, in the multivariate von Mises–Fisher distribution [33]

$$f\left(\mu\mid\mu_1,\kappa_1,\mu_2,\kappa_2\right)=c\left(\mu_1,\kappa_1,\mu_2,\kappa_2\right)\exp\left(\kappa_1\mu_1^T\mu+\kappa_2\mu_2^T\mu\right) \tag{3}$$

where $\mu_1$, and $\mu_2$ are the two preferred directions, and $\kappa_1$ and $\kappa_2$ denote the reliability parameters, that control bending. The factor $c\left(\mu_1,\kappa_1,\mu_2,\kappa_2\right)$ was used for normalization. Here, the preferred direction is determined based on the initial chain direction and the direction of the previous point. Therefore, $\kappa_1$ and $\kappa_2$ describe the reliability of the main orientation and last orientation of the fiber, respectively, and hence specify the bending smoothness. The cell aggregate packing applies a repulsion force to avoid crossing chains and a recovery force to ensure that the chain structure obtains a hard-core aggregate system [33]. The final formation of aggregates by the random walk approach is obtained as a balance between the repulsive and restorative forces. The actual aggregates model generation in this process will be stopped by a criterion in which the total force strength is reduced to a certain limit. The resulting cell chains can be used as a model of cyanobacterial aggregates for subsequent calculations.

Figure 2 shows the model realizations using input parameters for the quantity of cyanobacteria components, $N$ = 121, with an isotropic distribution of cyanobacteria orientation. The number of cyanobacteria objects in the aggregates was chosen based on the required volume fraction. Additionally, the realized volume fraction for each implementation is provided below.

(a)

(b)

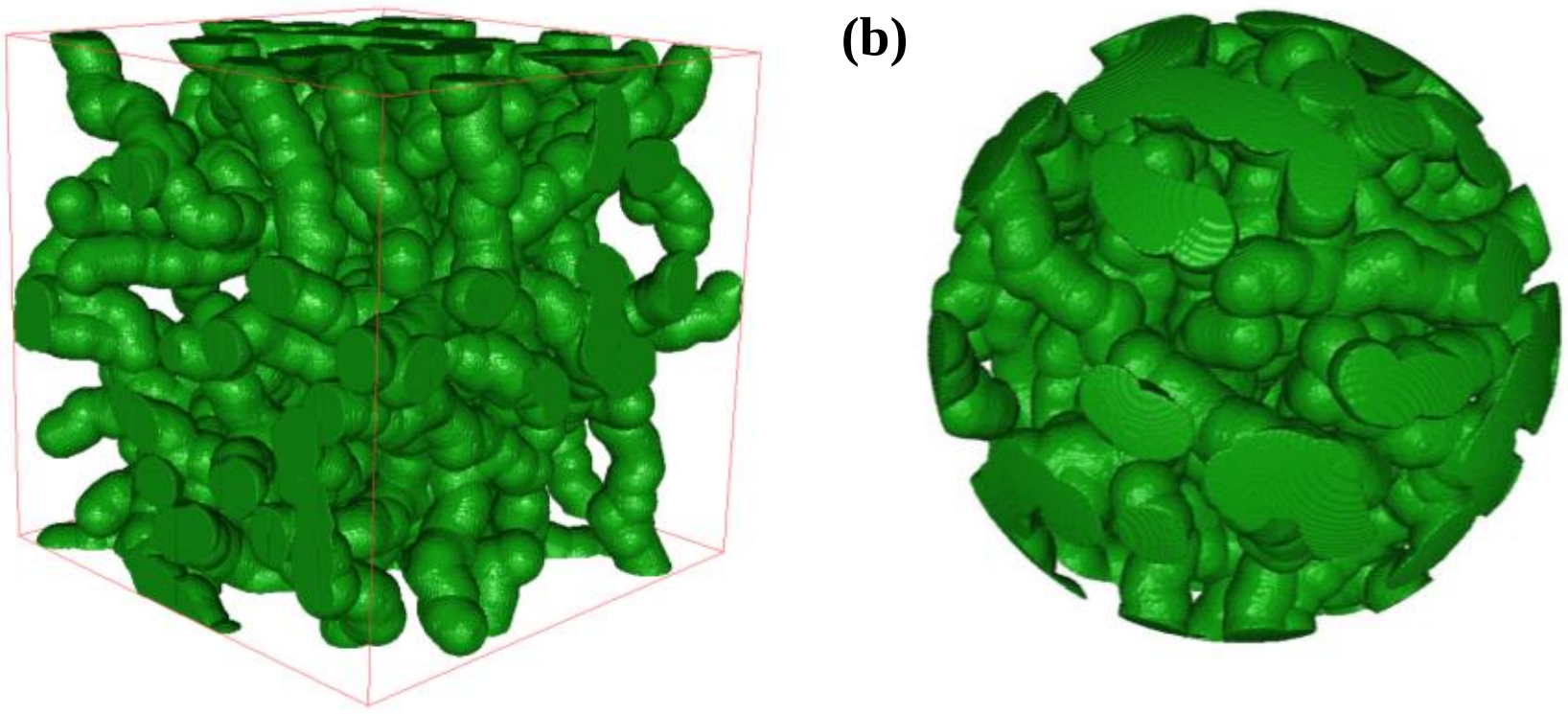

Fig. 2. Realizations for packed microalgae aggregates. The following are typical parameters: number of samples $n$ = 30 for each chain and bending parameters $\kappa_1$ = 5 and $\kappa_2$ = 5 are the dependability metrics in the desired directions. The volume fractions are (a) 37.9%, and (b) 44.8%.

The Gaussian random sphere's radius vector in spherical dimensions is defined as [35]

$$\mathbf{r}(\vartheta,\varphi)=\frac{a\exp\left[s(\vartheta,\varphi)\right]}{\sqrt{1+\sigma^2}}\mathbf{e}_r \tag{4}$$

$$s(\vartheta,\varphi)=\sum_{l=0}^{\infty}\sum_{m=-l}^{l}s_{lm}Y_{lm}(\vartheta,\varphi) \tag{5}$$

where $a$ and $\sigma$ are the mean and relative standard deviation of the radius vector, $(\vartheta,\varphi)$ are the spherical coordinates, $s$ is the so-called logradius, $\mathbf{e}_r$ a unit vector pointing outward in the direction $(\vartheta,\varphi)$, and $Y_{lm}$ the orthonormal spherical harmonics, and $s_{lm}$ the Gaussian random variables with zero means. The real and imaginary parts of a Gaussian random variable with zero mean and variance can be independently expressed as [35]

$$\mathrm{Var}\left[\mathrm{Re}\left(s_{lm}\right)\right]=\left(1+\delta_{m0}\right)\frac{2\pi}{2l+1}\ln\left(1+\sigma^2\right)C_l, \tag{6}$$

$$\mathrm{Var}\left[\mathrm{Im}\left(s_{lm}\right)\right]=\left(1-\delta_{m0}\right)\frac{2\pi}{2l+1}\ln\left(1+\sigma^2\right)C_l, \quad l=0,1,\ldots,\infty, \quad m=0,1,\ldots,l, \tag{7}$$

where $\delta_{m0}$ is the Dirac delta function, and $C_l$ is the nonnegative weight of degree $l$ in the Legendre polynomials. Moreover, the covariance function $\sum_s(\gamma)$, describes the auto-covariance of the random variables $s(\vartheta_1,\varphi_1)$ and $s(\vartheta_2,\varphi_2)$, which is given by a series of Legendre polynomials, [35]

$$\sum\nolimits_s(\gamma)=\ln\left(1+\sigma^2\right)\sum_{l=0}^{\infty}C_l P_l\left(\cos\gamma\right) \tag{8}$$

where $\gamma$ represents the angular distance between the two directions of $(\vartheta_1,\varphi_1)$ and $(\vartheta_2,\varphi_2)$. For the Legendre polynomials to be valid, it is required that

$$\sum_{l=0}^{\infty}C_l=1 \tag{9}$$

Practically, the infinite series representations in Eqs. (5) to (9) generally need to be truncated at some degree by $l_{\max}$. Thus, the Gaussian random sphere can be fully determined by the mean radius $a$ and the Legendre polynomial coefficient $C_l$. In addition, the conventional power-law covariance function for the representation $C_l$ can usually be used, [36]

$$C_l=\frac{\tilde{C}}{l^{\nu}}, \quad l=2,3,\ldots,l_{\max} \tag{10}$$

with $C_0 = C_1 = 0$, where $\tilde{C} = \left[\sum_{l=2}^{l_{\max}} l^{-\nu}\right]^{-1}$ is the normalization constant. Therefore, the Gaussian random sphere can be determined by the relative standard deviation $\sigma$, power-law index $\nu$, and the truncated degree $l_{\max}$. The power-law covariance function was chosen because it facilitates fine control of model size and translation variation, and it has been widely used to model natural cells of various shapes.

Figure 3 illustrates the Gaussian shaped microalgal aggregate system. The results are obtained by combining Gaussian random spheres and cyanobacterial aggregates of random walks. The Gaussian random sphere was generated using the same random sequence of numbers, therefore, the shapes' differences result from different weightings to the spherical harmonics coefficients $s_{lm}$. The random walks of the realization of the packed microalgae system were the same for the two different Gaussian sphere cases. Therefore, all the differences in the cases Figs. 3 (a) and (b) are mainly due to the external shape of the Gaussian sphere.

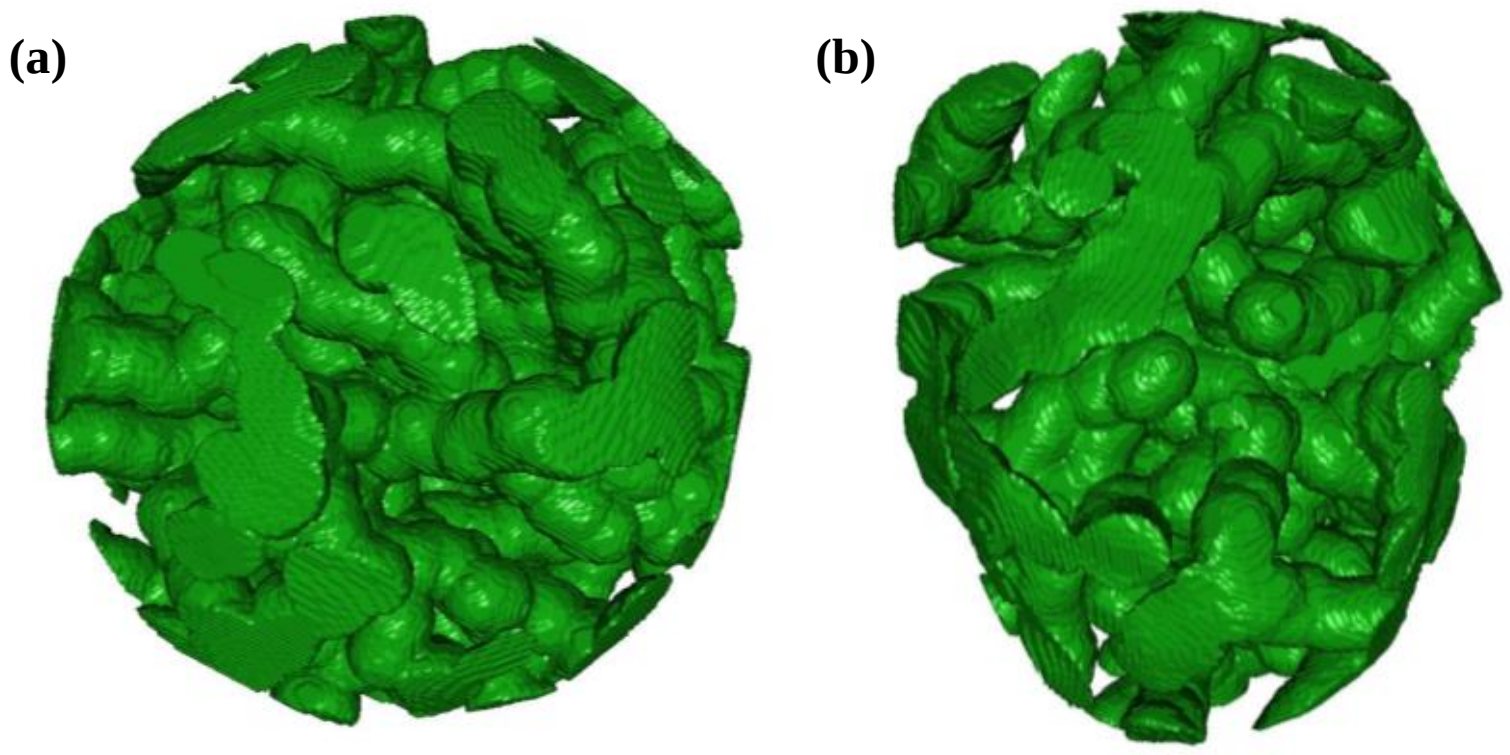

Fig. 3. Realizations for Gaussian spherical-shaped microalgae aggregates systems with power-law index $\nu$ = 3.5, truncated degree $l_{\max}$ = 50, and relative standard deviations of radial distance $\sigma$ =0.05, and 0.10 (from left to right). The common parameters of microalgae aggregates are the same as in the previous, and volume fractions are (a) 45.5%, and (b) 45.4%.

## 3. Scattering theory and methods

### 3.1. Lorenz-Mie theory review

Lorenz-Mie scattering is regarded as an important solution to Maxwell's equations for optically homogenous spheres. It is crucial to understand the scattering by spherical particles. Here, the external shape of the cyanobacterial aggregates was selected and considered to be approximately spherical. Therefore, the Lorenz-Mie scattering theory can be adopted to analyze and calculate the scattering/optical properties of microalgal aggregates. According to the electromagnetic scattering theory [37], the incident field, and scattered field can be expressed using vector spherical harmonics for a homogenous sphere. The coefficient of the external scattering field can be expressed as

$$a_n = \frac{m\psi_n(mx)\psi_n'(x) - \psi_n(x)\psi_n'(mx)}{m\psi_n(mx)\xi_n'(x) - \xi_n(x)\psi_n'(mx)} \tag{11}$$

$$b_n = \frac{\psi_n(mx)\psi_n'(x) - m\psi_n(x)\psi_n'(mx)}{\psi_n(mx)\xi_n'(x) - m\xi_n(x)\psi_n'(mx)} \tag{12}$$

and the expression for the coefficients of the internal electromagnetic field can be written as

$$c_n = \frac{m\psi_n(x)\xi_n'(x) - m\xi_n(x)\psi_n'(x)}{\psi_n(mx)\xi_n'(x) - m\xi_n(x)\psi_n'(mx)} \tag{13}$$

$$d_n = \frac{m\psi_n(x)\xi_n'(x) - m\xi_n(x)\psi_n'(mx)}{m\psi_n(mx)\xi_n'(x) - \xi_n(x)\psi_n'(mx)} \tag{14}$$

where $x$ is the size parameter, and $m$ represents the complex optical constants of the sphere medium. And the Riccati–Bessel functions are expressed as

$$\psi_n(x) = xj_n(x), \qquad \xi_n(x) = xh_n(x) \tag{15}$$

where $j_n(x)$ and $h_n(x)$ are spherical Bessel and Hankel functions, respectively. Observing the expressions for the coefficients of the internal and external electromagnetic fields, it can be seen that the denominators of $a_n$ and $d_n$ are identical, as well as the denominators of $b_n$ and $c_n$. This indicates that the internal field and scattered field modes are resonant for the same parameters of $x$ and $m$. This can be used to analyze and understand the relationship between the internal and external scattered electric fields. The scattering cross-section can be obtained by integrating the scattered power over all directions, and the extinction cross-section can be obtained using the optical theorem [37]

$$C_{sca} = \frac{2\pi}{k^2} \sum_{n=1}^{\infty} (2n+1)(|a_n|^2 + |b_n|^2) \tag{16}$$

$$C_{ext} = \frac{2\pi}{k^2} \sum_{n=1}^{\infty} (2n+1) \operatorname{Re}(a_n + b_n) \tag{17}$$

where $k$ denotes the wavenumber. The above formulas provide an understanding of individual particles' radiative/optical properties. We have reviewed the Lorenz-Mie theory above, although the calculation of the theory has been directly demonstrated once. The analytical solution of the Mie theory has high computational efficiency compared to other numerical methods, especially for large particles or microalgal aggregates. Moreover, the Lorenz-Mie theory is essential for understanding the scattering properties of cellular particles and their aggregates.

### 3.2. Scattering by particles

According to the electromagnetic theory, the incident and scattered fields can be related to the amplitude scattering matrix $\mathbf{S}$ for an arbitrary particle [37]

$$\mathbf{I}_s = \frac{1}{k^2 r^2} \mathbf{S}\,\mathbf{I}_i \tag{18}$$

where $\mathbf{I}_i$, and $\mathbf{I}_s$ represent the incident and scattering Stokes vector parameters of the incident light field, respectively. And $r$ represents the distance from the scatterer to observation points, $k$ is the wave number. The amplitude scattering matrix is block-diagonal with $S_{11}$, $S_{12} = S_{21}$, $S_{22}$, $S_{33}$, $S_{34} = -S_{43}$, and $S_{44}$ as non-zero elements for an arbitrary sphere, and the $S_{11}$ and $S_{22}$ are equal as are $S_{33}$ and $S_{44}$, which are a result of non-diagonal elements in the scattering amplitude matrix equal to zero [37]. The SPF $\Phi(\cos\Theta)$ can be directly obtained using the amplitude scattering matrix element $S_{11}$ [37]

$$\Phi(\cos\Theta) = \frac{4\pi S_{11}}{\mathrm{k}^2 C_{sca}} \tag{19}$$

This demonstrates how likely it is that light propagation in the solid angle $\mathrm{d}\Omega'$ around the direction $\mathbf{s}'$ will scatter into the solid angle $\mathrm{d}\Omega$ around the direction $\mathbf{s}$. For unpolarized incident waves, the scattering matrix ratio $-S_{12}/S_{11}$ represents the linear polarization degree of the scattered light. Moreover, the element ratio $S_{22}/S_{11}$, which is equal to 1 for a perfect sphere and describes the non-spherical nature of the scattering grains [38].

3.3. DDA method

The scattering properties of microalgal aggregates are studied by using the DDA method, which is a widely used approach to find field solutions for the scattering of arbitrarily shaped particles or aggregates [27, 39, 40]. The scatterer was divided into discrete dipoles to represent particles, or aggregates. The electric field values were solutioned from a group of linear equations based on the approximation. The solution of DDA for biological particles is more accurate owing to the relatively complex refractive index close to unity in water [41]. Therefore, the DDA method has innate strength for calculating the scattering properties of microalgal colonies. In our DDA calculations, the ADDA code released by Yurkin et al. was used [42, 43]. ADDA can run in parallel via MPI implementation, reducing the calculation time. Furthermore, Inzhevatkin and Yurkin [44] reported that the Wentzel-Kramers-Brillouin and its improved method as the initial guess in the DDA iterative solution process, which allowed us to accelerate the computational efficiency with the same accuracy. These features are beneficial to the calculation of the scattering properties of microalgal aggregates. Fig. 4 shows the scattering matrix elements of the cubical microalgae aggregate model (Fig. 2 a) for different dipoles per lambda (dpl) value. The refractive index is 1.021, and the absorption index is $1.12\times10^{-5}$. The matrix elements are orientation averaged over five different orientations. The size of the aggregate along the $x$ direction is 16.8 μm. Evidently, the elements converged for dpl = 6 compared with those for dpl = 12. The value of dpl was larger than 6 in the following simulation to ensure convergence. The scattering characteristics were orientation averaged over five different orientations for each realization. The lattice dispersion relation (LDR) for the polarizability of cubic dipoles was used in the calculations.

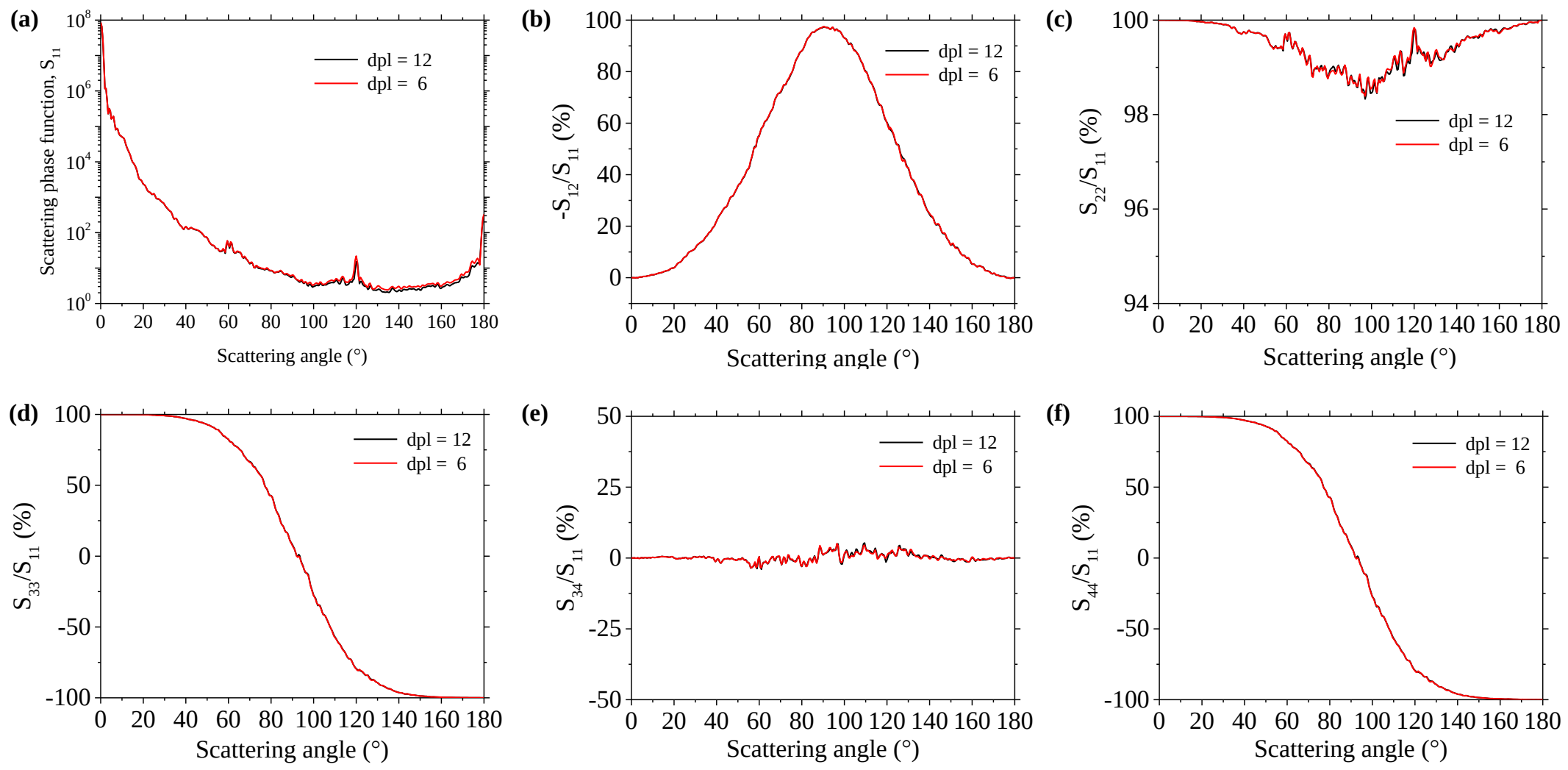

Fig. 4 Convergence verification of DDA solutions for dpl value.

## 3.4. Optical constants

The major metabolite compositions of microalgae include proteins, carbohydrates, and fats [45]. The effective complex optical constants of microalgae can be estimated from the effective medium approximation, for example, Maxwell-Garnett model [46], and Bruggeman's approach [47]. And the experimental refractive index of microalgae ranges from 1.004 to 1.167 because of the compositional variation of the components [19, 45, 48]. Moreover, the absorption index of the imaginary part of microalgae can be calculated according to the concentration of pigments in algae cell. Previous studies indicated that scattering is mainly influenced by the real part mismatch of the optical constants between the aggregates and the peripheral aqueous media, while absorption originates from the absorption index [24, 26, 49]. Therefore, variations in the refractive index indicate differences in scattering.

Moreover, the optical constants can be approximately considered a wavelength independent constant in the visible spectral range, as observed experimentally and theoretically [45, 48]. This indicates that the minor difference in the refractive index does not affect the conclusion of the scattering properties of microalgae. Here, the spectral dependent optical constants of microalgae was selected [19], to make the calculated results similar to the experimentally measured values. The spectral optical constants of the non-absorbing culture medium surrounding the cell can be determined by the Cauchy dispersion relation [28]

$$n_{\lambda} = A_1 + \frac{A_2}{\lambda^2} + \frac{A_3}{\lambda^4} \tag{20}$$

where the constants in the expression are $A_1$=1.32711, $A_2$ = 2.6 × 10$^{-3}$ μm$^2$, and $A_3$ = 5 × 10$^{-5}$ μm$^4$.

Figure 5 shows the spectral complex refractive index of microalgae cell, and the refractive index of culture medium. As shown, the refractive and absorption indices of the microalgae ranged from 1.344 to 1.351 and 2.118 × 10$^{-5}$ to 1.647 × 10$^{-3}$, respectively. The absorption peaks in the imaginary part of the complex optical constants correspond to *in vivo* chlorophyll a and chlorophyll b in the microalgae cells [50]. And the small dips in refraction index corresponding to the peaks, caused by Chlorophyll a, can be attributed to oscillator resonance theory [37]. The spectral optical constants of the culture medium decreased monotonically from 1.348 to 1.332.

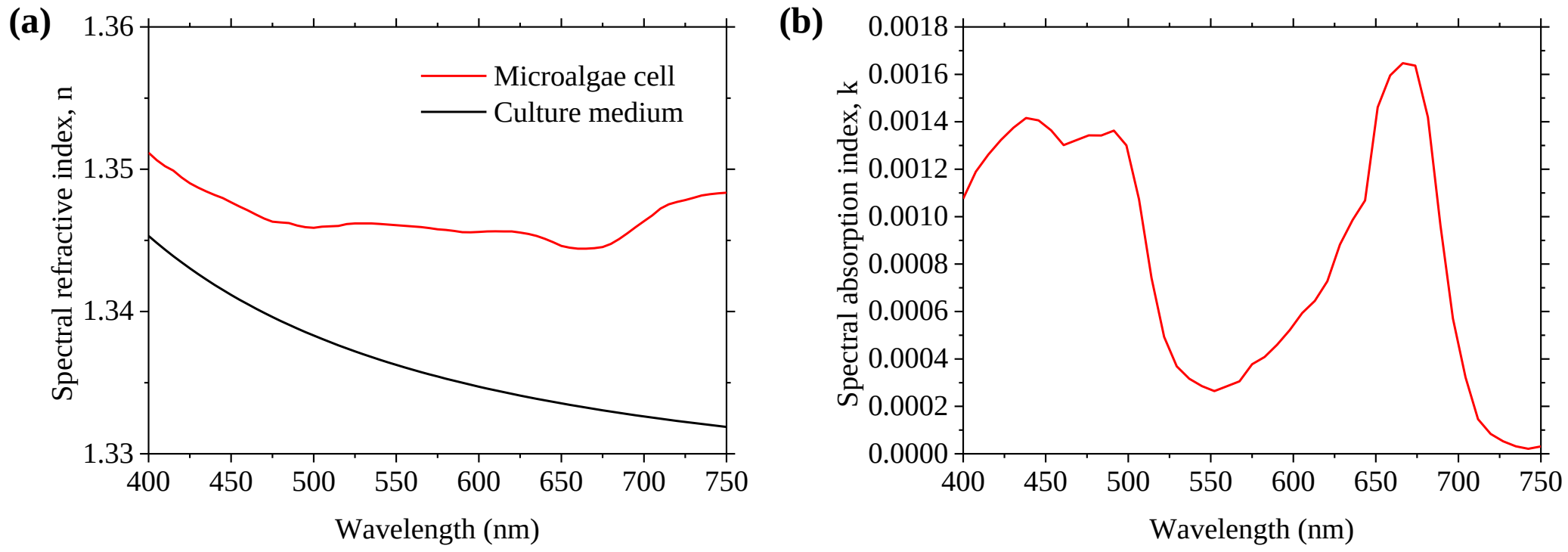

Fig. 5 The spectral complex refractive index of microalgae cell.

## 4. Results and discussion

4.1. Size and shape invariance of scattering elements

The scattering matrix elements could be employed in applications that consider the polarization of scattered radiation, such as remote sensing and phytoplankton identification. Fig. 6 shows the Mueller scattering matrix element ratios for spherical, cubic, and Gauss sphere shaped aggregates as functions of scattering angles using the DDA method. The same kind of Mueller matrix element ratios were predicted for 16.6 and 33.4 μm of microalgae aggregates for each shape. As shown, the Mueller matrix element ratios $-S_{12}/S_{11}$, $S_{33}/S_{11}$, and $S_{44}/S_{11}$ were identical for all shapes and sizes. The linear polarization degree of the algae aggregates $-S_{12}/S_{11}$ achieved 1 at 90° scattering angle, and was equal to 0 at 0° and 180° scattering angles. The Mueller matrix element ratios $S_{22}/S_{11}$, and $S_{34}/S_{11}$ were approaching for all aggregates and approximately equal to 1 and 0 for entire scattering angles, respectively. The Mueller matrix element ratio $S_{33}/S_{11}$ dropped from 1 to -1 when the scattering angle was changed from forward to backward direction, as well as the element ratio of $S_{44}/S_{11}$. The different Mueller matrix element ratios shown in Fig. 6 for microalgae aggregates are similar to those for a individual sphere [51]. Moreover, there were no resonance peaks appeared in the Mueller matrix element ratios of the multicellular filamentous cyanobacterial aggregates. These results are consistent with the result discovered by Mackowski and Mishchenko [52], which illustrated that aggregates caused the damping of oscillations in the matrix elements. And the disappearance of the oscillation was also attributed to the orientation averaging of the aggregates. Furthermore, the ratio $S_{22}/S_{11}$ deviated from unity was also observed by Mishchenko et al. [53], which can be used as an indicator of the nonsphericity of the aggregates.

The $S_{11}$ element exhibits a significant difference at the forward scattering angle, which increases as the aggregate size increases [54] (see Supplementary Materials). In addition to the increase in the forward direction of $S_{11}$, the scattering matrix elements can be approximately considered as shape and size invariant for microalgae aggregates.

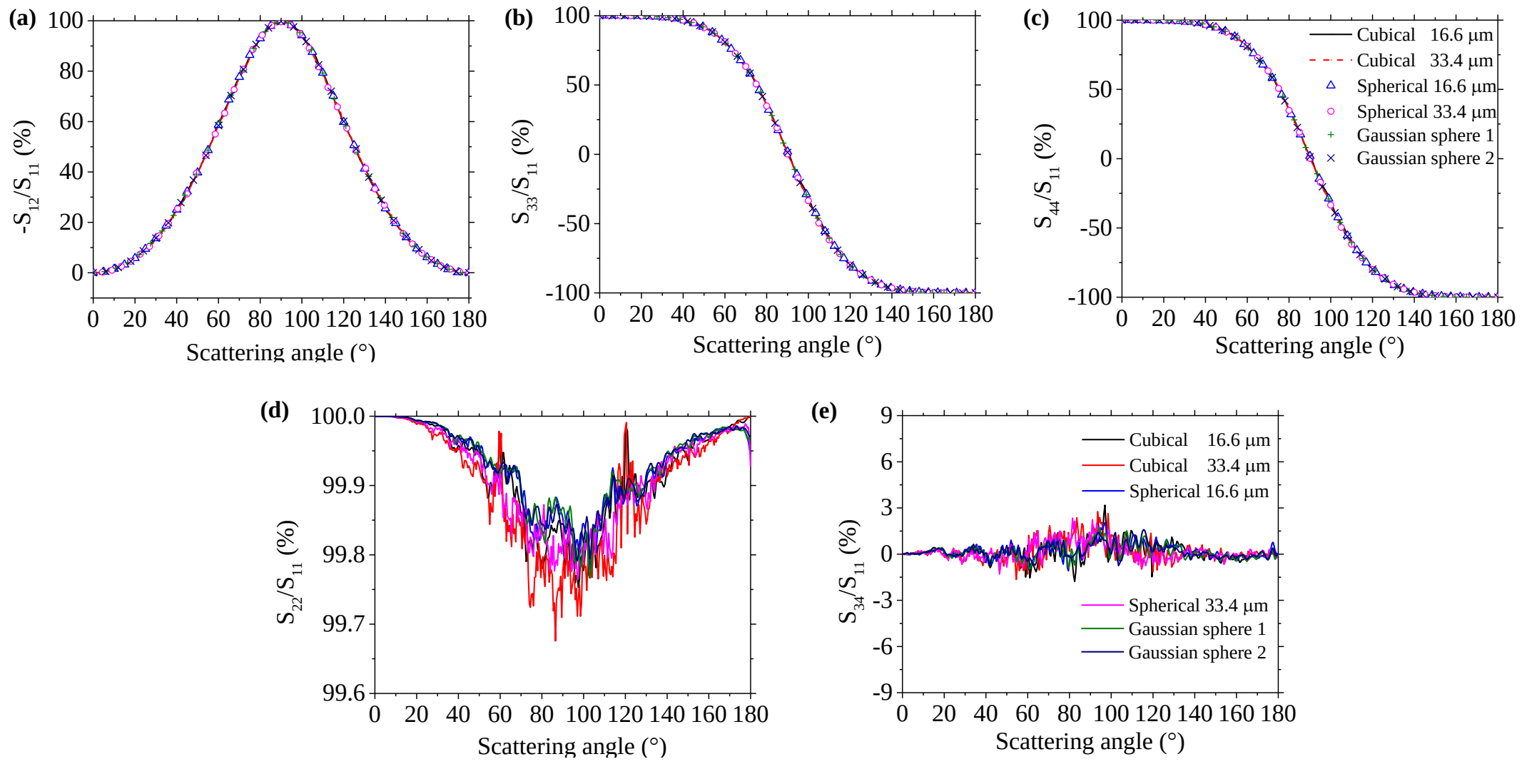

Fig. 6 The scattering matrix elements of spherical, cubical, and Gauss sphere-shaped aggregates for different sizes.

## 4.2. Size and shape invariance of cross-sections

Figure 7 demonstrates the EACSs per unit volume of the colonies of microalgae aggregates for different sizes, as 16.6, 33.4, and 66.5 μm. The cross-sections per unit volume of microalgae aggregates were defined as the cross-sections divided by the volume of cell aggregates. Fig. 7(a) indicates that the extinction cross-section presents an increasing trend with the increase in wavelength for all aggregate sizes. The extinction cross-section displays variations for different aggregate sizes, however, all variations lie in the interval of 0.08 to 0.18. Despite some differences, the extinction cross-sections were remarkably similar. The absorption cross-section almost overlaps in the spectral region of 510 to 650 nm and 680 to 750 nm, while the differences are relatively significant at absorption peaks around 450 nm. The non-overlapping of the cross-sectional parameters implies the presence of multiple scattering that cannot be ignored [55]. The absorption cross-sections corresponding to different sizes do not overlap, indicating that the absorption cross section is no longer a purely volumetric phenomenon [24, 25, 56]. This is attributed to the larger size of the aggregates and the absorption peaks in the absorption index, which are no longer simple values. The EACSs of the microalgae aggregates can be considered as approximately size invariant when overlooking some details. Moreover, the larger the microalgae aggregate size, the smaller the absorption cross-section, which indicates that an increase in the microalgae aggregate size will reduce light utilization efficiency.

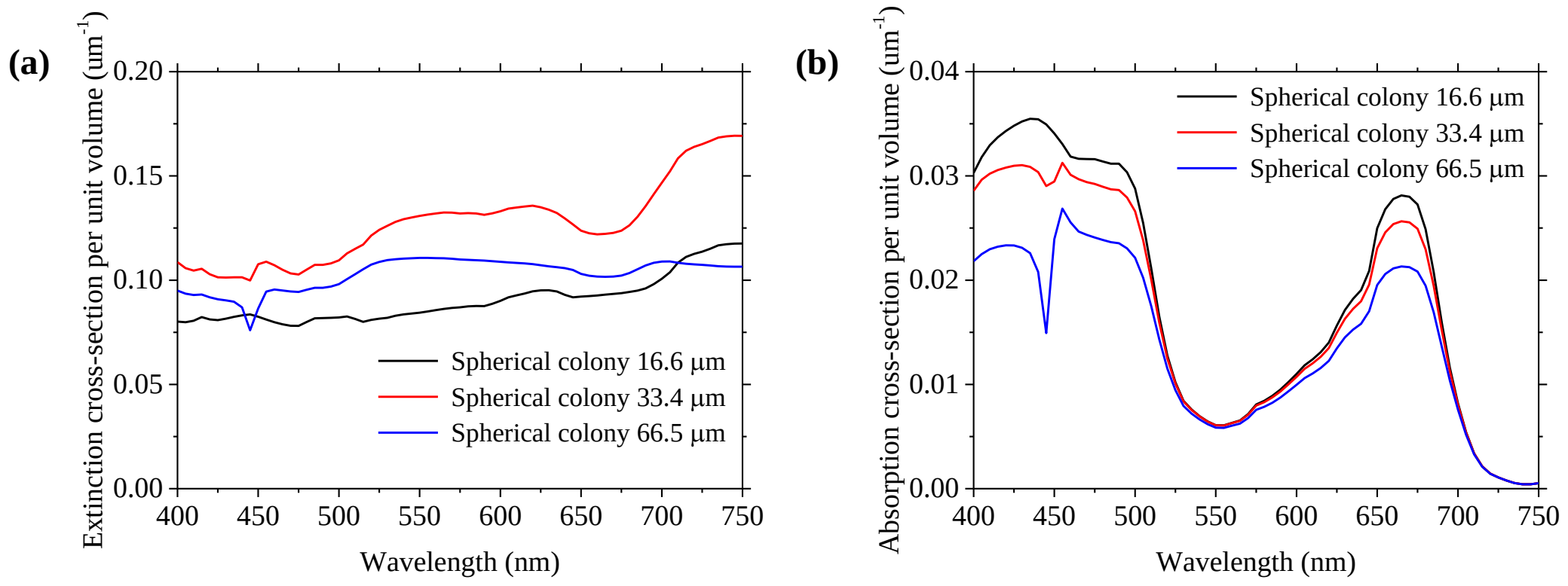

Fig. 7 The cross-sections per unit volume of the colonies of microalgae aggregates for different sizes.

Figure 8 shows the EACSs per unit volume of the colonies of microalgae aggregates for different shapes, i.e., spherical, cubical, and Gaussian. As shown, the variation trend of the extinction cross-section is consistent for different shapes and they nearly overlap over the PAR region. The absorption peaks of the absorption cross-sections of different shapes correspond completely and overlap perfectly in the PAR spectral region. This confirms on previous results reported by the authors [24, 25, 56], illustrating that absorption is a volumetric phenomenon. Therefore, the EACSs per unit volume are independent of the microalgal aggregate shape, which we call the shape invariance of the EACSs. This implies that there is no need to consider the aggregate shape's effect on the cross-sections' scattering properties.

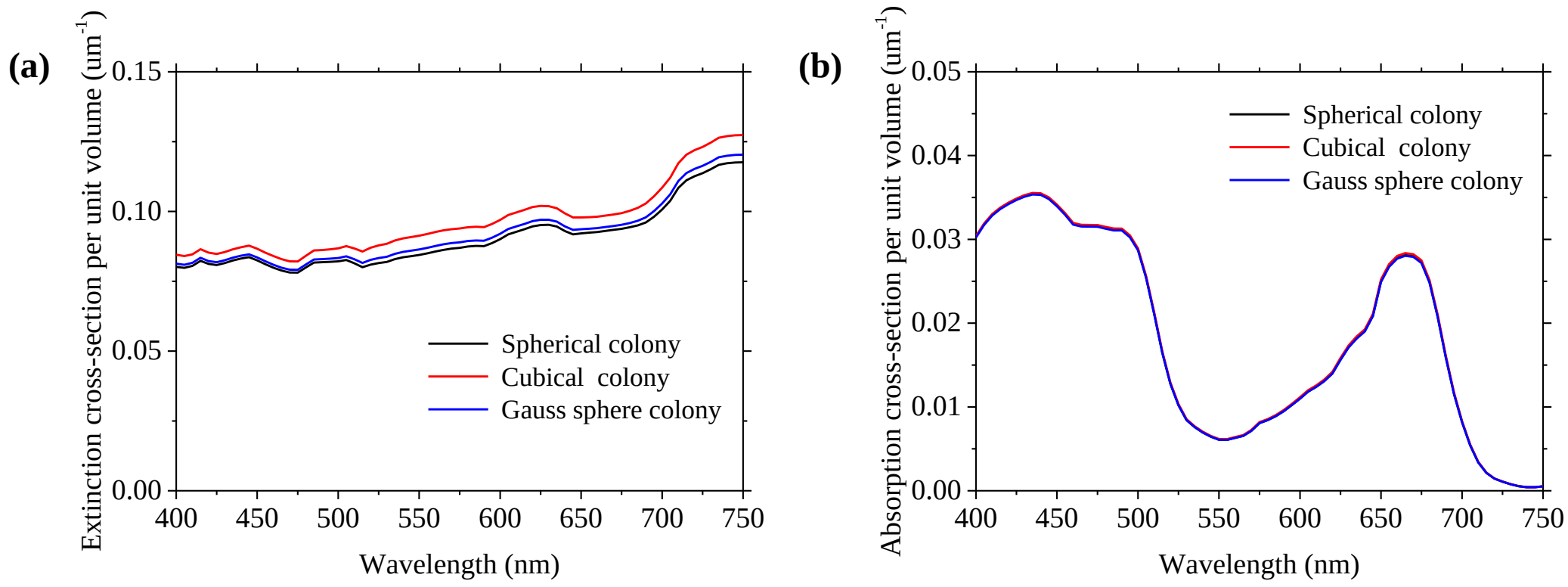

Fig. 8 The cross-sections per unit volume of the colonies of microalgae aggregates for different shapes.

Figure 9 presents the EACSs per unit volume of the colonies of microalgae aggregates with equivalent volumes for different shapes. As shown, the variation trend of the extinction cross-section was similar for different aggregate shapes, but they did not overlap over the PAR region. This can be attributed to extinction not being a volumetric phenomenon. Nevertheless, the absorption peaks of the absorption cross-sections correspond completely for different shapes, and nearly coincide in the PAR spectral region. Therefore, the absorption cross-sections can be considered independent of the volumetric distribution of the microalgae cells. This implies that the Lorenz-Mie theory can be employed to calculate the absorption cross-sections of microalgal aggregates under the volumetric phenomenon assumption.

To summarize, the EACSs per unit volume of the microalgal aggregates are shape and approximate size invariant. The absorption cross-section is not a fully volumetric phenomenon when the aggregate size is sufficiently large and the absorption peaks in the absorption index are considered.

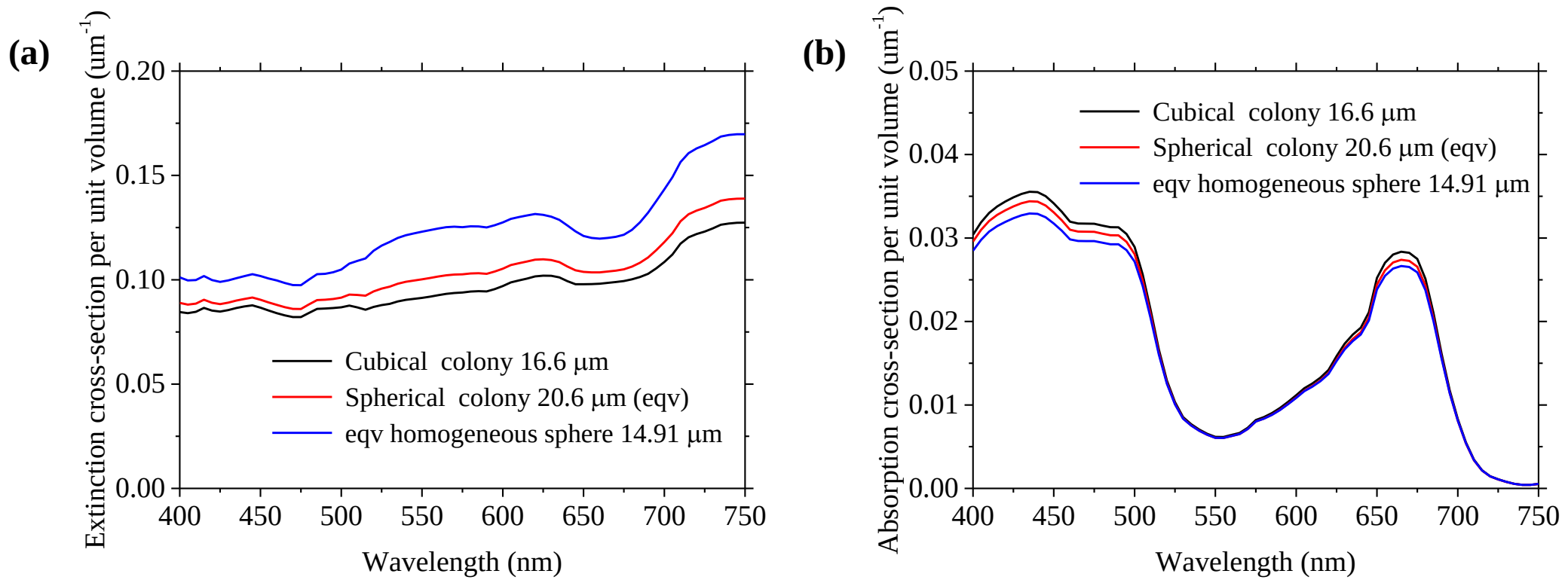

Fig. 9 The cross-sections per unit volume of the colonies of microalgae aggregates with equivalent volume.

### 4.3. Wavelength invariance of scattering matrix elements

As shown in Fig. 10, the scattering matrix elements of the colonies of spherically shaped microalgal aggregates. Previous section of this work validated that the Mueller matrix element ratios are independent of the shape and size of the aggregates. As shown, apart from $S_{11}$, the ratios of the scattering matrix element are also nearly wavelength invariant. However, element $S_{11}$ is proportional to the SPF, which can be approximately independent of wavelength in radiative transfer studies of microalgae [57]. The main difference in $S_{11}$ over the PAR wavelength is the forward scattering. And the microalgal asymmetry factor has been widely reported to be approximately 0.97 [13, 58].

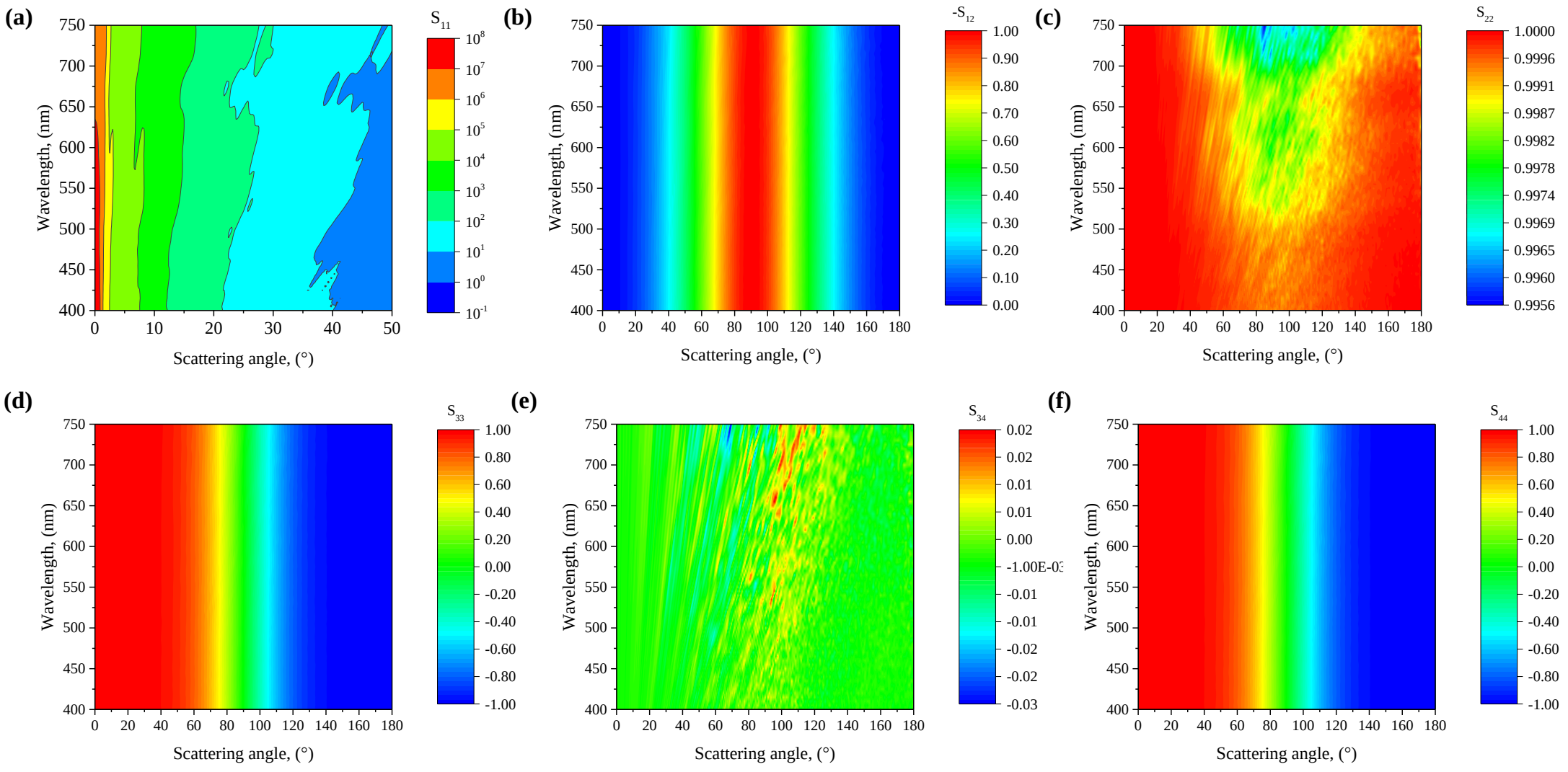

Fig. 10 The Mueller matrix elements of colonies of microalgae aggregates.

Furthermore, the light scattering in microalgae suspensions can be described by the H-G phase function to achieve reasonable accuracy [59]. The linear polarization degree of scattered electromagnetic field $-S_{12}/S_{11}$, which resembles a parabola that opens downwards, and equal to 0 in both the forward and backward orientations, and achieved 100% at the direction of 90° scattering angle for all wavelengths. The Mueller matrix element ratio $S_{22}/S_{11}$ was found to be nearly 100% for all scattering angles at the photosynthetically active radiation spectral range (equivalent to size parameters). The Mueller matrix element ratio $S_{34}/S_{11}$ can be considered to be approximately zero for all scattering angles and wavelengths. Moreover, the main diagonal element ratios in Fig. 10 (d) and (f) varied from 1 to -1 for the entire scattering directions, which can be considered identical for all wavelengths. These results are similar to those of individual spheres at all wavelengths, which due to the aggregates can be considered directionally independent. Combining the previously obtained conclusions on the invariance of the scattering element ratios with respect to shape and size, the scattering matrix element ratios are nearly shape, size, and wavelength invariant for microalgal aggregates.

4.4. Internal electric field of microalgae aggregates

The electromagnetic fields inside the aggregates were calculated in order to better understand and interpret the non-volumetric phenomena and large concavities exhibited by the absorption cross-section of the aggregates. Fig. 11 shows the real value of the total electric field of the colonies of spherical-shaped microalgae aggregates with a diameter of 66.5 μm on the $z = 0$ plane for the upper panel and $y = 0$ plane for the lower panel. The input parameters were the complex refractive indices at wavelengths of 440, 445, and 450 nm. As shown, the internal electric field of the aggregates at 445 nm shows a clear difference compared to that at 440 and 450 nm. Typical interference peaks at the periphery of the microalgal aggregates were observed. For fields at 445 nm, the internal field is relatively neglectable, except near the surface, which leads to a large dip in the absorption cross-section. This can be attributed to the resonance and non-resonance phenomena of the electromagnetic fields around the absorption peaks [60]. The presented internal field results confirm the cross-section oscillation results in the 440-450 nm spectral range, and explain the rapid decline and increase in the absorption cross-section between 440 and 450 nm.

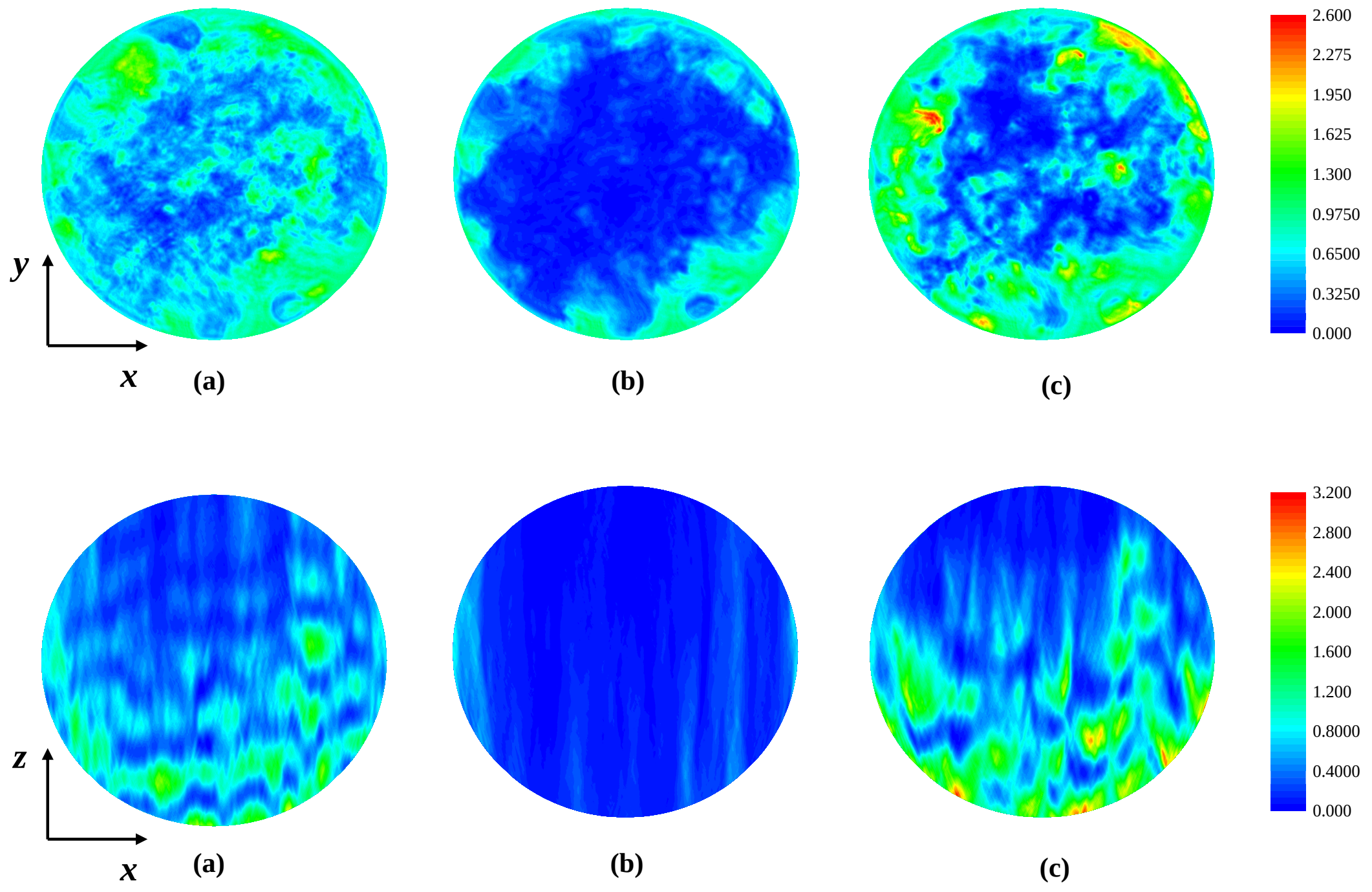

Fig. 11 The internal fields of colonies of microalgae aggregates, upper panel for $z = 0$ plane, and lower panel for $y = 0$ plane, the incident linearized polarized light at the propagation of $z$ direction.

## 5. Conclusions

On the basis of the results presented, rather than emphasizing the variation in scattering properties of microalgal aggregates with size, shape, and wavelength, this study explored on the possibility of invariance in the scattering properties. The scattering matrix element ratios of the cyanobacterial aggregates were nearly shape, size, and wavelength invariant. The EACSs per unit volume are shape invariant and approximately size invariant for cyanobacterial aggregates. And the absorption cross-section of the aggregates was independent of the spatial distribution of the microalgal cells. Furthermore, the absorption cross-section is not solely a volumetric phenomenon when the aggregates exceed a certain size. The invariant interpretation of the scattering properties of microalgal aggregates allows us to understand the scattering properties of cyanobacterial aggregates intuitively. The shape invariance of the scattering properties of the microalgal aggregates implies that the agglomeration process of algae and the motion of algal cells in the aggregates do not affect the light utilization of the cellular aggregates. The scattering characteristics of cyanobacterial aggregates are crucial for quantifying the interactions between light and photosynthetic microalgae.

**Funding**

The research was supported by Jiangxi Provincial Natural Science Foundation (No. 20212BAB214060) and China Scholarship Council, which are gratefully acknowledged.

**References**

[1] Jha D, Jain V, Sharma B, Kant A, Garlapati VK. Microalgae-based pharmaceuticals and nutraceuticals: an emerging field with immense market potential. ChemBioEng Reviews. 2017;4:257-72.

[2] Yap JK, Sankaran R, Chew KW, Halimatul Munawaroh HS, Ho SH, Rajesh Banu J, Show PL. Advancement of green technologies: A comprehensive review on the potential application of microalgae biomass. Chemosphere. 2021;281:130886.

[3] Pruvost J. Chapter 26 - Cultivation of Algae in Photobioreactors for Biodiesel Production. In: Pandey A, Larroche C, Dussap CG, Gnansounou E, Khanal SK, Ricke S, editors. Biofuels: Alternative Feedstocks and Conversion Processes for the Production of Liquid and Gaseous Biofuels (Second Edition): Academic Press; 2019.

p. 629-59.
[4] Revellame ED, Aguda R, Chistoserdov A, Fortela DL, Hernandez RA, Zappi ME. Microalgae cultivation for space exploration: Assessing the potential for a new generation of waste to human life-support system for long duration space travel and planetary human habitation. Algal Research. 2021;55:102258.
[5] Yang JH, Zhao T, Cui XY, Peng MB, Wang XT, Mao HM, Cui MS. New insights into the carbon neutrality of microalgae from culture to utilization: A critical review on the algae-based solid biofuels. Biomass and Bioenergy. 2022;166:106599.
[6] Yan N, Zhou K, Tong YW, Leong DT, Dickieson MP. Pathways to food from $CO_2$ via 'green chemical farming'. Nature Sustainability. 2022;5:907-9.
[7] Gifuni I, Pollio A, Safi C, Marzocchella A, Olivieri G. Current Bottlenecks and Challenges of the Microalgal Biorefinery. Trends in Biotechnology. 2019;37:242-52.
[8] Hoeniges J, Welch W, Pruvost J, Pilon L. A novel external reflecting raceway pond design for improved biomass productivity. Algal Research. 2022;65:102742.
[9] Pruvost J, Cornet JF, Le Borgne F, Goetz V, Legrand J. Theoretical investigation of microalgae culture in the light changing conditions of solar photobioreactor production and comparison with cyanobacteria. Algal Research. 2015;10:87-99.
[10] Galanakis CM. Microalgae: Cultivation, Recovery of Compounds and Applications: Academic Press; 2020.
[11] Muñoz R, Gonzalez-Fernandez C. Microalgae-based biofuels and bioproducts: from feedstock cultivation to end-products: Woodhead Publishing; 2017.
[12] Berberoglu H, Pilon L. Experimental measurements of the radiation characteristics

of Anabaena variabilis ATCC 29413-U and Rhodobacter sphaeroides ATCC 49419. International Journal of Hydrogen Energy. 2007;32:4772-85.

[13] Berberoglu H, Gomez PS, Pilon L. Radiation characteristics of Botryococcus braunii, Chlorococcum littorale, and Chlorella sp. used for $CO_2$ fixation and biofuel production. Journal of Quantitative Spectroscopy and Radiative Transfer. 2009;110:1879-93.

[14] Berberoglu H, Pilon L, Melis A. Radiation characteristics of Chlamydomonas reinhardtii CC125 and its truncated chlorophyll antenna transformants tla1, tlaX and tla1-CW+. International Journal of Hydrogen Energy. 2008;33:6467-83.

[15] Heng R-L, Pilon L. Time-dependent radiation characteristics of Nannochloropsis oculata during batch culture. Journal of Quantitative Spectroscopy and Radiative Transfer. 2014;144:154-63.

[16] Ma CY, Zhao JM, Liu LH. Experimental study of the temporal scaling characteristics of growth-dependent radiative properties of Spirulina platensis. Journal of Quantitative Spectroscopy and Radiative Transfer. 2018;217:453-8.

[17] Ma CY, Zhao JM, Liu LH, Zhang L. Growth-dependent radiative properties of Chlorella vulgaris and its influence on prediction of light fluence rate in photobioreactor. Journal of Applied Phycology. 2019;31:235-47.

[18] Zhao JM, Ma CY, Liu LH. Temporal scaling of the growth dependent optical properties of microalgae. Journal of Quantitative Spectroscopy and Radiative Transfer. 2018;214:61-70.

[19] Lee E, Heng R-L, Pilon L. Spectral optical properties of selected photosynthetic

microalgae producing biofuels. Journal of Quantitative Spectroscopy and Radiative Transfer. 2013;114:122-35.

[20] Pottier L, Pruvost J, Deremetz J, Cornet JF, Legrand J, Dussap CG. A fully predictive model for one-dimensional light attenuation by Chlamydomonas reinhardtii in a torus photobioreactor. Biotechnology & Bioengineering. 2005;91:569-82.

[21] Dauchet J, Blanco S, Cornet JF, Fournier R. Calculation of the radiative properties of photosynthetic microorganisms. Journal of Quantitative Spectroscopy & Radiative Transfer. 2015;161:60-84.

[22] Dong J, Zhao JM, Liu LH. Effect of spine-like surface structures on the radiative properties of microorganism. Journal of Quantitative Spectroscopy & Radiative Transfer. 2016;173:49-64.

[23] Bhowmik A, Pilon L. Can spherical eukaryotic microalgae cells be treated as optically homogeneous? Journal of the Optical Society of America A. 2016;33:1495-503.

[24] Heng RL, Sy KC, Pilon L. Absorption and scattering by bispheres, quadspheres, and circular rings of spheres and their equivalent coated spheres. Journal of the Optical Society of America A Optics Image Science & Vision. 2015;32:46-60.

[25] Kandilian R, Heng RL, Pilon L. Absorption and scattering by fractal aggregates and by their equivalent coated spheres. Journal of Quantitative Spectroscopy & Radiative Transfer. 2015;151:310-26.

[26] Lee E, Pilon L. Absorption and scattering by long and randomly oriented linear chains of spheres. Journal of the Optical Society of America A. 2013;30:1892-900.

[27] Ma CY, Zhao JM, Liu LH. Theoretical analysis of radiative properties of pronucleus multicellular cyanobacteria. Journal of Quantitative Spectroscopy and Radiative Transfer. 2019;224:91-102.

[28] Hoeniges J, Kandilian R, Zhang C, Pruvost J, Legrand J, Grizeau D, et al. Effect of colony formation on light absorption by Botryococcus braunii. Algal Research. 2020;50:101985.

[29] Harris EH. The Chlamydomonas Sourcebook: Introduction to Chlamydomonas and Its Laboratory Use: Academic press; 2009.

[30] Ratcliff WC, Herron MD, Howell K, Pentz JT, Rosenzweig F, Travisano M. Experimental evolution of an alternating uni- and multicellular life cycle in Chlamydomonas reinhardtii. Nature Communications. 2013;4:2742.

[31] Brennan L, Owende P. Biofuels from microalgae—A review of technologies for production, processing, and extractions of biofuels and co-products. Renewable and Sustainable Energy Reviews. 2010;14:557-77.

[32] Souliès A, Pruvost J, Legrand J, Castelain C, Burghelea TI. Rheological properties of suspensions of the green microalga Chlorella vulgaris at various volume fractions. Rheologica Acta. 2013;52:589-605.

[33] Altendorf H, Jeulin D. Random-walk-based stochastic modeling of three-dimensional fiber systems. Physical Review E. 2011;83:041804.

[34] Schladitz K, Peters S, Reinel-Bitzer D, Wiegmann A, Ohser J. Design of acoustic trim based on geometric modeling and flow simulation for non-woven. Computational Materials Science. 2006;38:56-66.

[35] Nousiainen T, McFarquhar GM. Light Scattering by Quasi-Spherical Ice Crystals. Journal of the Atmospheric Sciences. 2004;61:2229-48.

[36] Muinonen K, Zubko E, Tyynelä J, Shkuratov YG, Videen G. Light scattering by Gaussian random particles with discrete-dipole approximation. Journal of Quantitative Spectroscopy and Radiative Transfer. 2007;106:360-77.

[37] Bohren CF, Huffman DR. Absorption and scattering of light by small particles: John Wiley & Sons; 2008.

[38] Manickavasagam S, Mengüç MP. Scattering-matrix elements of coated infinite-length cylinders. Applied Optics. 1998;37:2473-82.

[39] Muinonen K, Pieniluoma T. Light scattering by Gaussian random ellipsoid particles: First results with discrete-dipole approximation. Journal of Quantitative Spectroscopy and Radiative Transfer. 2011;112:1747-52.

[40] Yurkin MA, Hoekstra AG. The discrete dipole approximation: An overview and recent developments. Journal of Quantitative Spectroscopy & Radiative Transfer. 2007;106:558-89.

[41] Draine BT, Flatau PJ. Discrete-Dipole Approximation For Scattering Calculations. Journal of the Optical Society of America A. 1994;11:1491-9.

[42] Yurkin MA, Hoekstra AG. The discrete-dipole-approximation code ADDA: Capabilities and known limitations. Journal of Quantitative Spectroscopy & Radiative Transfer. 2011;112:2234-47.

[43] Yurkin MA, Maltsev VP, Hoekstra AG. The discrete dipole approximation for simulation of light scattering by particles much larger than the wavelength. Journal of

Quantitative Spectroscopy and Radiative Transfer. 2007;106:546-57.

[44] Inzhevatkin KG, Yurkin MA. Uniform-over-size approximation of the internal fields for scatterers with low refractive-index contrast. Journal of Quantitative Spectroscopy and Radiative Transfer. 2022;277:107965.

[45] Aas E. Refractive index of phytoplankton derived from its metabolite composition. Journal of Plankton Research. 1996;18:2223-49.

[46] Mallet P, Gu, Eacute CA, Sentenac A. Maxwell-Garnett mixing rule in the presence of multiple scattering: Derivation and accuracy. Physical Review B. 2005;72:014205.

[47] Sihvola A. Mixing Rules with Complex Dielectric Coefficients. Subsurface Sensing Technologies & Applications. 2000;1:393-415.

[48] Jonasz M, Fournier GR. Light scattering by particles in water: Academic Press; 2011.

[49] Gordon HR. Light scattering and absorption by randomly-oriented cylinders: dependence on aspect ratio for refractive indices applicable for marine particles. Opt Express. 2011;19:4673-91.

[50] Bidigare RR, Ondrusek ME, Morrow JH, Kiefer DA. In-vivo absorption properties of algal pigments. International Society for Optics and Photonics. 1990;1302:290-303.

[51] Liu L, Mishchenko MI. Effects of aggregation on scattering and radiative properties of soot aerosols. Journal of Geophysical Research: Atmospheres. 2005;110.

[52] Mackowski DW, Mishchenko MI. Calculation of the T matrix and the scattering matrix for ensembles of spheres. Journal of the Optical Society of America A. 1996;13:2266-78.

[53] Mishchenko MI, Mackowski DW, Travis LD. Scattering of light by bispheres with touching and separated components. Applied Optics. 1995;34:4589-99.

[54] Manickavasagam S, Mengüç MP. Scattering matrix elements of fractal-like soot agglomerates. Applied Optics. 1997;36:1337-51.

[55] Yon J, Liu FS, Bescond A, Caumont-Prim C, Rozé C, Ouf FX, et al. Effects of multiple scattering on radiative properties of soot fractal aggregates. Journal of Quantitative Spectroscopy and Radiative Transfer. 2014;133:374-81.

[56] Drolen BL, Tien CL. Absorption and scattering of agglomerated soot particulate. Journal of Quantitative Spectroscopy and Radiative Transfer. 1987;37:433-48.

[57] Pilon L, Kandilian R. Chapter Two - Interaction Between Light and Photosynthetic Microorganisms. In: Legrand J, editor. Advances in Chemical Engineering: Academic Press; 2016. p. 107-49.

[58] Pilon L, Berberoğlu H, Kandilian R. Radiation transfer in photobiological carbon dioxide fixation and fuel production by microalgae. Journal of Quantitative Spectroscopy & Radiative Transfer. 2011;112:2639-60.

[59] Berberoglu H, Yin J, Pilon L. Light transfer in bubble sparged photobioreactors for $H_2$ production and $CO_2$ mitigation. International Journal of Hydrogen Energy. 2007;32:2273-85.

[60] Mishchenko MI, Hovenier JW, Travis LD. Light scattering by nonspherical particles: theory, measurements, and applications. New York: Academic Press; 1999.